\documentclass[12pt]{article}
\def\diy{\displaystyle}
\def\C{\mathbb C}

\def\P{\mathbb P}
\def\R{\mathbb R}

\def\Z{\mathbb Z}

\def\ffi{{\varphi}}

\def\wt{\widetilde}
\def\wh{\widehat}

\def\cA{{\rm A}}
\def\cB{{\rm B}}
\def\ccB{{\mathcal B}}

\def\cD{{\mathbb D}}

\def\cT{{\rm T}}
\def\cR{{\Sigma}}

\def\dist{{\rm{dist}}}
\def\diam{{\rm{diam}}}
\def\one{{\mathbf 1}}
\def\u0{{\mathbf 0}}

\def\uu{{\mathbf u}}
\def\uv{{\mathbf v}}

\def\ux{{\mathbf x}}
\def\uy{{\mathbf y}}

\def\ptt{{\widetilde{p}}}
\def\qtt{{\widetilde{q}}}
\def\rx{{{\rm x}}}
\def\ry{{{\rm y}}}
\def\ru{{{\rm u}}}
\def\rv{{{\rm v}}}

\def\wh{\widehat}
\def\wt{\widetilde}
\def\eps{{\epsilon}}
\def\om{{\omega}}
\def\Lam{{\Lambda}}
\def\Bphi{{\mbox{\boldmath${\phi}$}}}

\def\BLam{{\mbox{\boldmath${\Lam}$}}}

\def\Lamo#1{{\Lam_{#1}}}

\def\half{\frac{1}{2}}
\def\hhalf{1/2}
\def\pt{\partial}

\def\pmn{\par\medskip\noindent}
\def\psn{\par\smallskip\noindent}
\def\z2{{\Z^2}}
\def\zp2{{\Z^2_{\geq}}}

\def\myset#1{{\left\{\,#1\,\right\}}}

\def\pr#1{{  \P\left\{ \, #1 \, \right\}  }}

\def\dist{{\,{\rm dist}}}

\def\QED{{_{\;\;\blacksquare}}}

\def\myproof#1{{\pmn{\it Proof of #1. }}}

\def\truc#1#2#3{\smash{\mathop{\,\, #1 \,\, }\limits^{#2}_{#3}}}

\def\vbeta{{1/2}}

\def\valpha{{3/2}}

\def\DSzero{{\bf (DS.{\mbox{\boldmath${0, I}$}})}}

\def\DSk{{\bf (DS.{\mbox{\boldmath${k, I}$}})}}

\def\DSkone{{\bf (DS.{\mbox{\boldmath${k+1 ,I}$}})}}

\def\Szero{{\bf (S.{\mbox{\boldmath${0}$}})}}

\def\SSzero{{\bf (SS.{\mbox{\boldmath${0}$}})}}

\def\SSk{{\bf (SS.{\mbox{\boldmath${k}$}})}}

\def\SSkone{ {\bf (SS.{\mbox{\boldmath${k+1}$}})} }

\def\ISk{{\bf (IS.{\mbox{\boldmath${k}$}})}}

\def\ISkone{ {\bf (IS.{\mbox{\boldmath${k+1}$}})} }

\usepackage{graphics}

\usepackage{latexsym}
\usepackage{amsfonts, amssymb}
\usepackage{amsmath}
\usepackage{graphicx}
\usepackage{array}
\usepackage{rotating}

\newtheorem{Thm}{Theorem}[section]{\bf}{\it}
\newtheorem{Lem}{Lemma}[section]{\bf}{\it}
\newtheorem{Def}{Definition}[section]{\bf}{\it}
\newtheorem{Cor}{Corollary}[section]{\bf}{\it}

\numberwithin{equation}{section}

\begin{document}

\title{Eigenfunctions in a two-particle\\  Anderson
tight binding model}

\author{Victor Chulaevsky$^1$, Yuri Suhov$^2$ }                     
%

%
%



\maketitle
\pmn
$^1$ D\'{e}partement de Math\'{e}matiques et d'Informatique, \\
Universit\'{e} de Reims, Moulin de la Housse, B.P. 1039\\
51687 Reims Cedex 2, France
\psn
$^2$ Department of Pure Mathematics and Mathematical
Statistics\\ University of Cambridge \\
Wilberforce Road, Cambridge CB3 0WB, UK
\pmn
\pmn
\begin{abstract}
We establish the phenomenon of Anderson localisation for a quantum
two-particle system on a lattice $\Z^d$ with short-range interaction and
in presence of an IID external potential with sufficiently regular marginal
distribution.
\end{abstract}

\newpage

\section{The two-particle tight binding model.
\\Decay of Green's functions and localisation }
\label{intro}

This paper focuses on a two-particle Anderson tight binding model on lattice
$\Z^d$ with interaction. Our goal is three-fold. First, we establish a
theorem deducing exponential localisation
from a property of decay of Green's functions
in the two-particle model (Theorem \ref{thmone}
below). Second, we outline the so-called multi-scale analysis (MSA)
scheme for the two-particle model. Finally, we
perform the initial and the inductive steps of
two-particle MSA and therefore establish the
phenomenon of Anderson localisation in a two-particle
model for a large disorder. See our main result, Theorem 1.1.

We consider the Hilbert space of the two-particle system
$\ell_2(\Z^d\times\Z^d)$.
The Hamiltonian $H^{(2)}\left(=H^{(2)}_{U,V,g} (\om )\right)$
is a lattice Schr\"{o}dinger
operator of the form\\ $H^0+U+g(V_1+V_2)$, acting on functions
$\Bphi\in\ell_2(\Z^d\times\Z^d)$, given by
$$\begin{array}{ll}H^{(2)}\Bphi (\ux)&=H^0\Bphi (\ux)+
(U(\ux)+gW(\ux;\om))\Bphi(\ux )\\
\;&=\sum\limits_{\uy\in\Z^d\times\Z^d:\atop{\|\uy - \ux\|=1}}
\,\Bphi(\uy)
+  \left[U(\ux)+gW(\ux;\om)\right] \Bphi(\ux),\\
W(\ux;\om) &= V(x_1;\om) + V(x_2;\om),\quad\ux=(x_1,x_2)
\in\Z^{d}\times\Z^{d}.
\end{array}\eqno(1.1)$$
Here and below we use boldface letters such as $\ux, \uy, \uu$ etc.
for points in $\Z^d \times \Z^d$.
Next,  $x_j=\big(\rx_j^{(1)},\ldots,\rx_j^{(d)}\big)$ and
$y_j=\big(\ry_j^{(1)},\ldots, \ry_j^{(d)}\big)$ stand for the
coordinate vectors of particles in $\Z^d$,
$j=1,2$, and $\|\cdot\|$ is the sup-norm: for
$\uv = (v_1,v_2)\in\R^d\times\R^d$:
$$\| \uv\| =\max_{j=1,2}\; \|v_j\|,\eqno (1.2.1)$$
where
$$\|v\|= \max_{i=1, \dots, d}\left|{\rm v}^{(i)}\right|, \text{ for }
v=({ \rm v}^{(1)}, \ldots, { \rm v}^{(d)})\in\R^d.\eqno (1.2.2)$$
We will consider the distance on $\R^d\times\R^d$ and $\R^d$
generated by the norm $\|\;\cdot\;\|$.

Throughout this paper, the random external potential
$V(x;\omega )$, $x\in\Z^d$, is assumed to be
real IID, with a common distribution function ${\rm F}_V$ on $\R$.
Of course, the random variables $W(\ux;\om)$ form an array with
dependencies (which is the main source of difficulties in spectral analysis
of multi-particle quantum systems in random environment).
\pmn

A popular assumption is that ${\rm F}_V$ has a probability density
function (PDF).
The condition on ${\rm F}_V$ guaranteeing the validity of all results
presented in
this paper is as follows:
$$\begin{array}{l}\hbox{\sl ${\rm F}_V$ has a {\rm{PDF}}
${\rm p}_V$ }\hbox{\sl which is bounded and has a compact support.}
\end{array}\eqno (1.3)
$$
This will allow us to use, in Section 3,
some results on single-particle localisation
proved in \cite{A94} with the help of the fractional-moment
method (FMM), an alternative of the MSA for single-particle
models; see \cite{AM}, \cite{ASFH}). We note that a number
of important facts proven or used here
remain true under considerably weaker assumptions on ${\rm F}_V$.
For example, Wegner-type bounds (3.5), (3.6) hold under the
condition that for some $\delta>0$ and all $\eps>0$,
$$
\sup_{a\in \R}\;({\rm F}_V(a+\eps)-{\rm F}_V(a))\leq\eps^{\delta};
$$
see \cite{CS1}. Moreover, we stress that with the help
of a technically more
elaborate argument it is possible to obtain the main result of
this paper (Theorem \ref{MThm} below) under an assumption weaker
than (1.3). We would also like to note that the IID property of
$\{V(x;\om), x\in\Z^d\}$ can be relaxed.
See our forthcoming manuscript \cite{CS2}.

Parameter $g\in\R$ is traditionally called the coupling, or
amplitude, constant.

The interaction potential $U$ is assumed to satisfy the
following properties.
\psn
(i) $U$ {\sl is a bounded
real function $\Z^{d}\times\Z^d\to\R$ symmetric
under the permutation of variables: $U({\mathbf x})=
U(\sigma{\mathbf x})$, where}
$$
\sigma {\mathbf x}=(x_2, x_1)
\;\hbox{ for }\; {\mathbf x}=(x_1,x_2), \; x_1, x_2 \in\Z^d.
\eqno(1.4)
$$
(ii) $U$ {\sl obeys}
$$U(\ux)=0,\;\hbox{ if }\;\|x_1-x_2\| > r_0,
\eqno(1.5)
$$
Here $r_0\in [1,+\infty)$ is a given value (the interaction range).

Let $\P$ stand for the joint probability distribution of RVs
$\{V(x;\om), x\in\Z^d\}$. The main assertion of this paper is

\begin{Thm} \label{MThm} Consider the two-particle random Hamiltonian
$H^{(2)}(\om)$ given by {\rm{(1.1)}}.
Suppose that $U$ satisfies conditions {\rm{(1.4)}} and {\rm{(1.5)}},
and the random potential
$\{V(x;\om)$, $x\in\Z^d\}$ is {\rm{IID}},  with a marginal distribution
function ${\rm F}_V$ obeying {\rm{(1.3)}}. Then there exists
$g^*\in(0,+\infty)$ such that for any $g$ with $|g|\geq g^*$,
with $\P$-probability one, the spectrum of operator $H^{(2)}(\om )$
is pure point. Furthermore, there exists a nonrandom constant
$m_+ = m_+(g) >0$ (the effective mass) such that all eigenfunctions
$\Psi_j(\ux;\om)$ of $H^{(2)}(\om )$ admit an exponential bound:
$$
|\Psi_j(\ux;\om)|\leq C_j(\om) \, e^{-m_+\|\ux\|}.\eqno(1.6)$$
\end{Thm}

The assertion of Theorem \ref{MThm} can also be stated in the form
where $\forall$ given $m_*>0$, $\exists$ $g_*=g_*(m_*)\in (0,+\infty )$
such that $\forall$ $g$ with $|g|\geq g_*$, the eigenfunctions
$\Psi_j(\ux;\om)$ of $H^{(2)}(\om )$ admit exponential bound (1.6)
with effective mass $m_+\geq m_*$.

The conditions of Theorem \ref{MThm} are assumed throughout the paper.
As was said earlier, the proof of Theorem \ref{MThm}  uses mainly MSA,
in its two-particle version. Most of the time we will work with
finite-volume approximation operators $H^{(2)}_{\BLam_{L}(\uu)}\left(=
H^{(2)}_{\BLam_{L}(\uu)} (\om )\right)$ given by
$$
H^{(2)}_{\BLam_{L}(\uu)} =
H^{(2)}\upharpoonright_{\BLam_L(\uu)}
+ \text{ Dirichlet boundary conditions }
\eqno(1.7)
$$
and acting on vectors $\Bphi\in\C^{\BLam_{L}(\uu)}$ by
$$
\begin{array}{r}H^{(2)}_{\BLam_L(\uu)}\Bphi (\ux)=
\sum\limits_{\uy\in\BLam_L(\uu ):\atop{\|\uy - \ux\|=1}}
\,\Bphi(\uy)
+  \left[U(\ux)+gW(\ux;\om)\right] \Bphi(\ux),\quad\quad\\
\ux=(x_1,x_2)\in{\BLam_L(\uu)},
\end{array}
\eqno(1.8)
$$
with $W(\ux)$ as in (1.1). Here and below,
$\BLam_L(\uu)$ stands for the `two-particle lattice box' (a box,
for short) of size $2L$ around
$\uu=(u_1, u_2)$, where $u_j=(\ru^{(1)}_j, \ldots,
\ru^{(d)}_j)\in\Z^d$:
$$
\BLam_L(\uu) = \left(
\truc{\times }{2}{j=1} \truc{\times }{d}{i=1}
[\ru_j^{(i)}-L, \ru_j^{(i)}+L ]\right)\cap\left(\Z^d\times\Z^d
\right).\eqno (1.9)$$
Denoting by $\left|\BLam_L(\uu )\right|$ the cardinality of
$\BLam_L(\uu )$, $H^{(2)}_{\BLam_L(\uu)}$ is a Hermitian operator
in the Hilbert space $\ell_2(\BLam_L(\uu))$ of dimension
$\left|\BLam_L(\uu )\right|$.

In fact, the approximation (1.7) can be used for any finite
subset $\BLam\subset\Z^d\times\Z^d$ of cardinality $|\BLam |$
producing Hermitian operator $H^{(2)}_{\BLam}$ in $\ell_2(\BLam)$.

Hamiltonian $H^{(2)}$ and its approximants $H^{(2)}_\BLam$
admit the permutation symmetry.
Namely, let $S$ be the unitary operator in
$\ell_2(\Z^d\times\Z^d)$ induced by map $\sigma$:
$$
S \Bphi (\ux) = \Bphi(\sigma\ux ).
\eqno(1.10)$$
Then $S^{-1}H^{(2)}S =H^{(2)}$ and $S^{-1}H^{(2)}_\BLam S
=H^{(2)}_{\sigma\BLam}$ (with natural
embeddings $\C^{\BLam},\C^{\sigma\BLam}\subset\ell_2(\Z^d\times\Z^d)$).
This implies, in particular, that for
any finite $\BLam \subset \Z^d\times\Z^d$, the eigenvalues
of operators $H^{(2)}_\BLam$ and $H^{(2)}_{\sigma\BLam}$ are identical.
This fact is accounted for in the course of presentation.
\pmn

Like its single-particle counterpart, the two-particle
MSA scheme involves a number of technical parameters
playing roles similar to those in the paper \cite{DK}. In this
and the following section we make use of
some of these parameters (to begin with, see Theorem \ref{thmone}).
More precisely, given a positive number $\alpha>1$ and
starting with $L_0>0$ large enough and $m_0>0$,
define an increasing sequence $L_k$:
$$
L_k = L_0^{\alpha^k},\;\; k\geq 1,
\eqno(1.11)
$$
and a decreasing positive sequence $m_k$ (depending on a positive number $\gamma$):
$$
m_k = m_0 \,\prod_{j=1}^k \left( 1 - \gamma L_k^{-1/2} \right)
,\;\; k\geq 1.
\eqno (1.12)
$$
We will also use in Theorem \ref{thmone} parameter $p$; our assumptions on
$\alpha$, $\gamma$ and $p$ in this theorem will be that
$$
p>\alpha d >1,\;\;\gamma \geq 40.\eqno(1.13)
$$
Note that sequence $m_k$ is indeed positive, and the
limit $\lim\limits_{k\to\infty} m_k  \geq m_0/2$ when
$L_0$ is sufficiently large. We will also assume that $L_0>r_0$.

The single-particle MSA scheme was used in \cite{DK} to check,
for IID potentials, decay properties of the Green's functions
(GFs). In this paper we adopt a similar strategy. For the
two-particle model, the GFs in a box $\BLam_L(\uu )$ are
defined by:
$$
G^{(2)}_{\BLam_L(\uu)}(E; \ux, \uy) =\left\langle \left(
H^{(2)}_{\BLam_L(\uu)} - E\right)^{-1}
\delta_{\ux}, \delta_{\uy}
\right\rangle, \; \ux,\uy\in\BLam_L(\uu),
\eqno(1.14)
$$
where $\delta_{\ux}(\uv )= \one \big(\uv =\ux\big)$ is the lattice
Dirac delta-function (considered as a vector in $\C^{\BLam_L (\uu)}$).
Following \cite{DK}, we introduce

\begin{Def} Fix $E\in \R$ and $m>0$. A two-particle box
$\BLam_L(\uu)$ is said to be $(E,m)$-non-singular
(in short: $(E,m)${\rm{-NS}}) if the GFs
$G^{(2)}_{\BLam_L(\uu)}(E;\uu,\uu')$ defined by
(1.14) for the Hamiltonian
$H^{(2)}_{\BLam_L(\uu)}$ from (1.8) satisfy
$$\truc{\max}{}{\uy\in\pt \BLam_L(\uu)}
\left| G^{(2)}_{\BLam_L(\uu)}(E;\uu,\uy) \right| \leq e^{-m L}.
\eqno(1.15)$$
\psn
Otherwise, it is called $(E,m)$-singular (or $(E,m)${\rm{-S}}).
Here $\pt \BLam_L(\uu)$ stands for the interior boundary
(or briefly, the boundary) of box $\BLam_L(\uu)$: it
is formed by points $\uy\in\BLam_L(\uu)$ such that $\exists$ a site
$\uv\in\big(\Z^d\times\Z^d\big)\setminus\BLam_L(\uu)$ with $\|\uy -\uv\|=1$.
A similar concept can be introduced for any set $\BLam\subset\Z^d\times\Z^d$,
for which we use the same notation $\pt\BLam$.
\end{Def}

The first step in the proof of Theorem \ref{MThm}  is Theorem \ref{thmone} below.
More precisely, Theorem \ref{thmone} deduces exponential localisation from a postulated
property of decay of two-particle GFs.
The proof of Theorem \ref{thmone} given in Section 2
follows that of its
single-particle counterpart from earlier works (see Section 1 in
\cite{FMSS} and Theorem 2.3 from \cite{DK}). Nevertheless,
this theorem is an important part of our method (as in \cite{FMSS}
and \cite{DK}); having established Theorem \ref{thmone} one can
attempt to prove two-particle localisation by analysing only GFs
$G^{(2)}_{\BLam_L(\uu)}(E;\uu,\uy)$.
\pmn

It is convenient to introduce the following

\begin{Def}
A pair of two-particle boxes $\BLam_L(\uu)$, $\BLam_L(\uv)$
is called $R$-distant ($R$-D, for short) if
$$
\min\left\{ \|\uu -  \uv\|, \|\sigma \uu -  \uv\| \right\} > 8R.
\eqno(1.16)
$$
Here, $\sigma$ was defined in (1.4).
\end{Def}

\begin{Thm}\label{thmone}
Let $I\subseteq \R$ be an interval. Assume that for
some $m_0>0$ and $L_0>1$,
$\lim\limits_{k\to\infty} m_k  \geq m_0/2$,
and for any $k\geq 0$ the following properties hold:
$$
\DSk \;\;\;
\begin{array}{l}\hbox{$\forall$ $\uu, \uv \in \Z^d\times\Z^d$
such that $\Lam_{L_k}(\uu)$ and $\Lam_{L_k}(\uv)$ are
$8L_k$-D,} \\
\pr{\forall \,E\in I:\;
\;\Lam_{L_k}(\uu)\;{\rm{or}}\;\Lam_{L_k}(\uv)\;
{\rm{is}}\;(m_k,E){\rm{-NS}}}\geq 1 - L_k^{-2p}.\end{array}
\eqno(1.17)
$$
Here $L_k$ and $m_k$ are defined in {\rm{(1.11)}},  {\rm{(1.12)}},
and $\sigma$ by {\rm{(1.4)}},
with $p$, $\alpha$ and $\gamma$ satisfying  {\rm{(1.13)}}. Then,
with probability one, the spectrum of operator $H^{(2)}(\om)$ in
$I$ is pure point. Furthermore, there exists a constant
$m_+\geq m_0/2$ such that all
eigenfunctions
$\Psi_j(\ux;\om)$ of $H^{(2)}(\om)$ with eigenvalues
$E_j(\om)\in I$ decay exponentially fast at infinity,
with the effective mass $m_+$:
$$
|\Psi_j(\ux;\om)|\leq C_j(\om) \, e^{-m_+\|\ux\|}.
\eqno(1.18)
$$
\end{Thm}

We stress that it is the property \DSk $\,$
encapsulating decay of the
GFs which enables the two-particle MSA scheme to work. (Here
and below, DS stands for `double singularity').
\psn

Clearly, Theorem 1.1 would be proved, once the validity of
property \DSk $\,$
had been established
for $I=\R$ and for all $k\geq 0$. However,
an important remark is that, to deduce Theorem \ref{MThm},
we actually need to check the conditions of Theorem \ref{thmone}
for an arbitrary interval $I\subset\R$ of unit length (but of course
with a fixed sequence
of values $m_k$ and $L_k$ from (1.11), (1.12)). In fact, by covering
the whole spectral line $\R$ by a countable family of such intervals,
we will get that the whole spectrum of $H^{(2)}$ is pure point with
$\P$-probability one, with a `universal' effective mass $m_+>0$.

We will therefore focus on establishing property \DSk $\,$
for an arbitrary unit interval $I$ and all $k\geq 0$; this is
done in Sections 3--5 below. Nevertheless, many details of the
presentation in Sections 3-5
do not require the assumption that the length of $I$ is $1$;
we will choose appropriate conditions on an {\it ad hoc} basis.

\section{Proof of Theorem \ref{thmone}}

It is well-known (see, e.g., \cite{B}, \cite{S}) that
almost every energy $E$
with respect to the spectral measure of $H^{(2)}$ is a generalised
eigenvalue of $H^{(2)}$, i.e., solutions $\Psi$ of the equation
$H^{(2)}\Psi = E \Psi$ are polynomially bounded.
Therefore, it suffices to prove that the generalised eigenfunctions
of $H^{(2)}$ decay exponentially with $\P$-probability one.

Let $E\in I$ be a generalised eigenvalue of Hamiltonian
$H^{(2)}$ from Eqn (1.1), and $\Psi$ be a corresponding generalised
eigenfunction. Following \cite{DK}, we will prove that
$$
\forall {\wt\rho}\in (0,1):\;\;{\limsup_{\|\ux\|\to\infty}} \,\,\ln
\frac{ \left| \Psi(\ux;\om) \right|  }{\|\ux\|}
\leq - {\wt\rho} \, m,
\eqno(2.1)
$$
where $m>0$ is the constant from the statement of Theorem 1.1.

Given $\uu\in\Z^d\times\Z^d$ and an
integer $k=0, 1, 2, \ldots$, set
$$
R(\uu) = \|\sigma\uu - \uu \|, \;b_k(\uu )= 1 + R(\uu) L_{k}^{-1},\;
{\mathbf M}_{k}(\uu ) = \BLam_{L_{k}}(\uu) \cup \sigma\BLam_{L_{k}}(\uu);
\eqno (2.2)$$
cf. (1.4). Note that $\forall$ $\uu\in\Z^d\times\Z^d$,
$$
\forall\;k\geq 1:\;
{\mathbf M}_{k}(\uu) \subset \BLam_{b_{k} L_{k}}(\uu),
\;\hbox{ and $\lim\limits_{k\to\infty}b_k(\uu )= 1$.}\eqno (2.3)
$$

Now set
$$
{\mathbf A}_{k+1}(\uu) = \BLam_{b_{k+1}L_{k+1}}(\uu) \setminus
\BLam_{b_{k} L_{k}}(\uu)\eqno (2.4)
$$
and define the event
$$
\Omega_k(\uu)
= \{ \exists\, E\in  I\text{ and }\ux\in {\mathbf A}_{k+1}(\uu):
\, \BLam_{L_{k}}(\ux) \text{ and }  \BLam_{L_{k}}(\uu) \text{ are }
(m,E)-{\rm S}\}.\eqno (2.5)
$$
Observe that, owing to the definition of ${\mathbf M}_{L_{k+1}}(\uu)$
(see (2.2)), if
$\ux\in {\mathbf A}_{k+1}(\uu)$, then
$$
\dist\left( \BLam_{L_{k}}(\ux), \, \left[  \BLam_{L_{k}}(\uu) \cup
\sigma\BLam_{L_{k}}(\uu)\right]
\right) \geq 8 L_{k}.
$$
Thus, by the hypothesis of the theorem,
$$
\pr{ \Omega_k(\uu)} \leq \frac{ (2b_{k+1}L_{k+1} + 1)^{2d}}{ L_k^{2p}}
\leq \frac{ (2b_{k+1} + 1)^{2d}}{ L_k^{2p-2\alpha }}.
\eqno(2.6)
$$
Since $p>\alpha$ and by virtue (2.3), the series
$$
\sum_{k=0}^\infty \pr{\Omega_k(\uu)}<\infty.\eqno(2.7)
$$

Consider the event
$$
\Omega_{<\infty}(\uu) =
\{  \forall \, \uu\in\Z^d\times\Z^d, \, \Omega_k(\uu)
\text{ occurs finitely many times}\}.\eqno(2.8)
$$
Then, owing to (2.7) and the Borel--Cantelli lemma,
$\pr{\Omega_{<\infty}}=1$.
So, it suffices to pick a potential sample $\{V(y;\om),y\in\Z^d\}$
with $\om\in\Omega_{<\infty}$ and prove
exponential decay of any generalised eigenfunction $\Psi$
of operator $H^{(2)}$, with the respective eigenvalue $E\in I$,
for the specified $\om\in\Omega_{<\infty}$. From now on, the
argument in the proof of Theorem becomes
deterministic, and we omit symbol $\om$, except for the places
where its presence is instructive.

Since $\Psi$ is polynomially bounded, there exist $C,t\in (0,+\infty )$
such that
$$
\forall\, \ux\in\Z^d\times\Z^d, \,\, |\Psi(\ux) | \leq C \, (1+\|\ux\|)^t.
\eqno (2.9)$$
Further, since $\Psi$ is not identically zero, $\exists$
$\uu\in\Z^d\times\Z^d$ such that
$\Psi(\uu) \neq 0$. For any given $k$, if
$E\not\in {\rm{spec}}\left(H^{(2)}_{\BLam_{L_k}(\uu)}\right)$ then
the values of $\Psi$ inside $\BLam_{L_k}(\uu)$
can be recovered from its boundary values, in particular,
$$
\Psi(\uu) = \sum_{\uv\in \pt \BLam_{L_k}(\uu)}\;\sum_{\uv'
\in\pt^+\BLam_{L_k}(\uu)}{\mathbf 1}_{\{\|\uv-\uv'\|=1\}}
G^{(2)}_{\BLam_{L_k}(\uu)}(E; \uu, \uv) \, \Psi(\uv').
\eqno(2.10)
$$
Here and below, $\pt^+\BLam_{L_k}(\uu)$ denotes the exterior
boundary of box $\BLam_{L_k}(\uu)$ formed by the points
${\mathbf y}\in\pt\left(\Z^d\times\Z^d\setminus\BLam_{L_k}(\uu)\right)$,
i.e., the points ${\mathbf y}\in\Z^d\times\Z^d\setminus\BLam_{L_k}(\uu)$
with dist $({\mathbf y},\BLam_{L_k}(\uu))=1$.

Suppose that a two-particle box $\BLam_{L_k}(\uu)$ is $(m,E)$-NS
for an infinite number of values $k$
(i.e. for arbitrarily big values of $k$). Then, by Definition 1.1,
$$
\left|G^{(2)}_{\BLam_{L_k}(\uu)}(E; \uu, \uv)\right|
\leq e^{-m L_k},
$$
yielding, by (2.9) and (2.10), that
$$
| \Psi(\uu)|
\leq 4d(2L_k+1)^{2d-1} \,e^{-m L_k}  C(1 + \|\uu\| + L_k)^t \to 0,
\;{\rm{as}}\;k\to\infty.\eqno (2.11)
$$
This would mean that, in fact, $\Psi(\uu)=0$, which contradicts
the above assumption.
Therefore, $\exists$ an integer $k_1 = k_1(\om, E, \uu)
<\infty$ such that
$\forall\, k\geq k_1$ box $\BLam_{L_k}(\uu)$ is $(m,E)$-S. At
the same time, owing to
the choice of $\om\in\Omega_{<\infty}$, $\exists\, k_2 =k_2(\om,\uu)$
such that $\forall\, k\geq
k_2$, the event $\Omega_k(\uu)$ does {\it not } occur.
Then
$$\forall\, k\geq \max\big\{k_1,k_2\big\}:\;
\forall\, \ux\in{\mathbf A}_{k+1}(\uu),\,
\text{ box }
\BLam_{L_k}(\ux) \text{ is  $(m,E)$-NS }.
\eqno (2.12)
$$
Next, for  a given $\rho\in(0,1)$ and $b> \frac{1 + \rho}{1-\rho}$, define
$$
{\wt{\mathbf A}}_{k+1}(\uu)
= \BLam_{2b/(1+\rho)L_k}(\uu) \setminus \BLam_{2/(1-\rho)L_k}(\uu)
\subset{\mathbf A}_{k+1}(\uu).
$$
Naturally, the above inclusion is valid for sufficiently large $L_{k+1}$, or, with $L_0>1$ being fixed, for
sufficiently large values of $k$, which we will assume in this argument.

For any $\ux\in {\wt{\mathbf A}}_{k+1}(\uu)$, we have that
$$
\dist (\ux, \pt A_{k+1}(\uu)) \geq \rho \, \| \ux - \uu \|.
$$
Furthermore, if $\| \ux - \uu \| \geq L_0/(1 - \rho)$, then
$\exists \, k\geq 0$ such that $\ux\in {\wt{\mathbf A}}_{k+1}(\uu)$.

For any $k\geq\max\big\{k_1,k_2\big\}$, box $\BLam_{L_k}(\uu)$ must
be $(m,E)$-NS, and therefore, $E\not\in{\rm{spec}} \left(H^{(2)}_{\BLam_{L_k}(\uu)}\right)$. Hence, Eqn (2.10) holds,
with $\ux$ instead of $\uu$:
$$
\Psi(\ux) = \sum_{\uv\in \pt \BLam_{L_k}(\ux)}\;\sum_{\uv'
\in\pt^+\BLam_{L_k}(\ux)}{\mathbf 1}(\|\uv-\uv'\|=1)
G^{(2)}_{\BLam_{L_k}(\ux)}(E; \uu, \uv) \, \Psi(\uv').
\eqno(2.13)
$$
Further, by virtue of non-singularity of $\BLam_{L_k}(\uu)$, $\exists \,
\uv_1\in\pt^+ \BLam_{L_k}(\uu)$ such that
$$
| \Psi(\ux)|
\leq 4d(2L_k+1)^{2d-1} \, \,e^{-m L_k}\, | \Psi(\uv_1)|.\eqno (2.14)
$$
In fact, it suffices to apply bound (1.15) and pick, in the RHS of
(2.13), a point $\uv'$ incident to $\uv$
providing the maximal absolute value of the GF
$G^{(2)}_{\BLam_{L_k}(\ux)}(E; \uu, \uv)$.

Next, pick a value ${\wt\rho}\in (0,1)$ and write it
as a product
${\wt\rho} = \rho \rho'$, where
$\rho, \rho'\in (0,1)$. Further, pick any
$b > 8 + 1 + {\rho}/(1-\rho)$. We can iterate bound (2.14)
at least $((L_k +1)^{-1}\rho \|\ux - \uu\|$
times, obtaining the following inequality:
$$
|\Psi(\ux)|
\leq \left( 4d(2L_k+1)^{2d-1} \, e^{-mL_k}
\right)^{(L_k+1)^{-1}\rho\|\ux - \uu\|}
\, C \left(1 + \|\uu\| + bL_{k+1} \right)^t.
\eqno(2.15)
$$
This provides an exponential decay rate of $\Psi$ arbitrarily
close to $\rho$. Namely, $\exists$ an integer $k_3 \geq
\max\big\{k_1,k_2\big\}$ such that $\forall$
$k\geq k_3$, if $\|\ux - \uu\| \geq L_k/(1-\rho)$ then
$$
|\Psi(\ux)| \leq e^{- \rho \rho' m \|\ux - \uu\|}.$$
Therefore,
$$
{\limsup_{\|\ux\|\to\infty} } \,\,\frac{1}{ \|\ux\|} \,\ln | \Psi(\ux)|
\leq - \rho \rho' m.$$
Eqn (2.1) then follows. This completes the proof of Theorem
\ref{thmone}. $\quad\QED$
\psn

\section{The two-particle MSA scheme.\\ Non-interactive
pairs of singular  boxes}

In view of Theorem \ref{thmone}, our aim is to check property \DSk $\,$
in Eqn (1.17).
We now outline the two-particle MSA which is used for this
purpose. It bears many features
borrowed from its single-particle counterpart. In both
single- and two-particle versions,
the MSA scheme is an elaborate induction dealing with GFs
$G^{(2)}_{\BLam_{L_k}(\uu )}$ and
involving several mutually related parameters;  some of them have
been used in Sections 1 and 2. Here we give the complete list,
following specifications (1.11), (1.12) of sequences $L_k$ and $m_k$
(these specifications are assumed for the rest of the paper).

$\bullet$ Parameter $\alpha\in(1,2)$
determines the rate of growth of sequence $L_k$ in Eqn (1.11).

$\bullet$ Constant $\gamma>0$  determines
sequence $m_k$ in Eqn (1.12).

$\bullet$ Parameter  $p>0$ controls the power of the decay
of probability of double singularity; see property \DSk $\,$
in (1.17).

 $\bullet$ Parameters $m_0>0$ (the initial mass) and
 $L_0>1$ (the initial length) define the initial step of the
induction. These values are related to the threshold
$g^*\in(0,\infty)$ for the coupling constant $g$ in Theorem \ref{MThm},
roughly, by the constraint $m_0L_0 \sim \ln g^*$; see the proof
of the initial inductive step in Theorem \ref{MSAInd0} below. In addition,
to complete the inductive step, $L_0$
should be large enough: $L_0\geq L^*$; see Theorem \ref{MSAInd}.

$\bullet$ Parameters $m_k>0$ and $L_k>1$
(the mass and the length at step $k$) are chosen to
follow Eqns (1.11) and (1.12) since it allows us to
check Eqn (1.17) with substantial use of the
single-patricle MSA scheme.

Next,

$\bullet$ Parameter  $\beta\in(0,1)$ controls
the important property of tolerated resonances;
see Eqn (3.2).

$\bullet$ Parameter $q>0$ is responsible for decay
of probability of non-tolerated resonances.

Thresholds  $L^*$ and $g^*$ are functionally described as
$L^*=L^*(d, \beta, \alpha , m_0, p, q)$ and $g^*=g^*(d, \beta, \alpha,
m_0, p, q)$. An initial insight into the values of these thresholds
is provided by writing
$L^*=\max\;[L^*_0,L^*_1]$ and $g^*=\max\;[g^*_0,g^*_1]$. Here
$L^*_0$ and $g^*_0$ are (rather explicitly) determined
from Eqn (3.1) and Theorem \ref{MSAInd0} whereas
$L^*_1$ and $g^*_1$ are encrypted into Theorem \ref{MSAInd}.
A further insight into the values of $L^*$ and $g^*$ is provided
in the course of the presentation below. The
initial mass $m_0>0$ can be chosen at will (but of course the choice
of $m_0$ affects that of $L^*$ and $g^*$).

To start with, we assume
$L_0\geq L^*_0$ where $L^*_0$ is large enough, so that
$$
\prod_{j=1}^\infty \left( 1 - \gamma L_j^{-1/2} \right)
\geq \half\,.\eqno (3.1)
$$
Technically, it is convenient for us to run the two-particle MSA
scheme under the following conditions on parameter values:
$$
p >12d+9,\, q > 4p + 12d,\, \beta = \vbeta, \, \alpha = \valpha,
\gamma=40.
\eqno(3.2)
$$
We will assume Eqns (3.1), (3.2) for the rest of the paper,
although we will use symbols $\alpha$ and $\beta$ to make
analogies with \cite{DK} more fulfilling.

We will also use results of the single-particle MSA formulated
and proved in \cite{DK}.
The property of decay (with high probability) of GFs of the
single-particle Anderson
tight binding model is proved, in particular, under assumption
of large disorder:
$|g|\geq \tilde g>0$, where the threshold $\tilde g$ is defined
in terms of the single-particle
Hamiltonian. We always assume, directly or indirectly, that the
two-particle threshold $g^*$,
introduced in this paper, satisfies $g^*\geq \tilde g$, so that
for all $g$ with $|g|\geq g^*$,
all results of \cite{DK} for the single-particle
model are valid. In order to avoid confusion,
we will denote by
$\ptt$ and $\qtt$ parameters analogous to $p$ and $q$ but
related to
the single-particle model.
It is worth mentioning that, according to the results of \cite{DK},
one can choose
$\ptt$ and $\qtt$ arbitrarily large, provided that $|g|$ is
sufficiently large. Therefore,
we can also assume that $\ptt$ and $\qtt$ are as large as required
for our arguments, provided that
$\tilde g$ is sufficiently large and
$$
|g|\geq g^*\geq \tilde g
\eqno(3.3)
$$

The initial step of the two-particle MSA scheme
consists in establishing properties
\Szero $\,$
and \SSzero;
see Eqns (3.7) and (3.8).
The inductive step of the two-particle MSA
consists in deducing property \SSkone $\,$
from property \SSk; again see Eqn (3.8).
These properties are equivalent
to properties \DSzero,
\DSkone $\,$
and \DSk, $\,$
respectively,
figuring in Theorem \ref{thmone} (in the form of
{\bf (SS.\,{\mbox{\boldmath${\;\cdot\;}$}})} they are slightly
more convenient to deal with).
Both the initial and the inductive step are done with tghe assistance
of properties {\bf (W1)} and/or {\bf (W2)} (Wegner-type
bounds, see Eqns (3.5) and 3.6) below) which should be established
independently. In our context,
i.e. for a two-particle system,
properties {\bf (W1)} and {\bf (W2)} have been proved in \cite{CS1}.
\pmn

\begin{Def}\label{}
Given $E\in\R$, $\uv\in\Z^d\times\Z^d$ and
 $L>1$, we call the box $\BLam_L(\uv)$
$E$-resonant (briefly: $E$-R) if the spectrum
of the Hamiltonian
$H^{(2)}_{\BLam_L(\uv )}$ satisfies
$$
{\rm dist}\left[E, {\rm{spec}}\left(H^{(2)}_{\BLam_{L}(\uv )}
\right)\right]< e^{-L^\beta}.\eqno(3.4)
$$
\end{Def}

Given an $L_0>1$,
introduce the following properties {\bf (W1)} and
{\bf (W2)} of Hamiltonians $H^{(2)}_{\BLam_l}$,
$l\geq L_0$.
$$\hbox{{\bf (W1)} $\quad$
$\forall$ $l\geq L_0$, box $\BLam_l(\ux)$ and $E\in\R$: $\;$}
\pr{ \BLam_l(\ux) \text{ is } E\text{\rm-R} }< l^{-q}.
\qquad\;\;\;\eqno (3.5)$$

$$\hbox{\bf (W2) $\quad$}\begin{array}{l}\hbox{
$\forall \, l\,\geq L_0$  \hbox{ and $8 l$-D boxes $\BLam_\ell(\ux)$
and $\BLam_\ell(\uy)$,}}\\ \;
\pr{ \exists \, E\in\R:\;{\rm{both}}\;\BLam_l(\ux) \text{ and }
\BLam_l(\uy) \text{ are } E{\rm{-R}} } < l^{-q}.\end{array}\quad\;\;
\eqno (3.6)$$

\begin{Lem}\label{WW}{\rm{(Cf. \cite{CS1}.)}}
Under the above assumptions on $\{V(x;\om)\}$ and $U$
(see (1.3)-(1.5)), properties ${\bf W1, W2}$ hold true.
\end{Lem}

Further, let $I\subseteq\R$ be an interval. Given
$m_0>0$ and $L_0>1$, consider property \Szero$\,$:

$$\Szero \quad \begin{array}{l}\hbox{
$\forall$ $\ux\in\Z^d$,}
\;\;\pr{\exists \, E\in I:\;\;\BLam_{L_0}(\ux)
\;{\rm{is}}\;(E,m_0){\rm -S} }<L_0^{-2p}. \quad\quad\quad
\end{array}\eqno (3.7)$$

Next, for interval $I\subseteq\R$ and
values $L_k$ and $m_k$, $k\geq 0$, as in {\rm{(1.11)}}
and {\rm{(1.12)}}, we introduce property \SSk:
$$
\SSk\quad
\begin{array}{l}
\hbox{$\forall$ $L_k${\rm -D}
boxes $\BLam_{L_k}(\ux)$ and $\BLam_{L_k}(\uy)$:}\\
\;\;\pr{\exists \, E\in I:
\;{\rm{both}}\;\BLam_{L_k}(\ux),\BLam_{L_k}(\uy)
\;{\rm{are}}\;(E,m_k){\rm -S} } <L_k^{-2p}.
\end{array}
\eqno (3.8)$$

The initial MSA step is summarised in
\pmn

\begin{Thm}\label{MSAInd0} $\forall$ given $m_0$ and $L_0>0$
and $\forall$ bounded interval $I\subset\R$, there exists
$g^*_0=g^*_0(m_0,L_0,|I|)\in (0,+\infty)$ such that
for $|g|\geq g^*_0$, properties \Szero $\,$ and \SSzero $\,$ hold true.
\end{Thm}
\pmn

\myproof{Theorem {\rm{\ref{MSAInd0}}}}
Obviously, property \Szero$\;$ implies \SSzero, so we
focus on the former. Property
\Szero$\;$ is established along
the lines of \cite{DK}; see \cite{DK}, Proposition A.1.2. Without loss
of generality, we can assume that $g>0$.
Let $E_0\in\R$ be the middle point of $I$ and $2\eta$ be its length:
$I=(E_0-\eta,E_0+\eta)$. Note that  if
$\forall\, \ux=(x_1,x_2)\in \BLam_{L_0}(\uu)$ we have
$$
 \left| W(\ux) - E_0  \right|  \geq 4d + 2\eta + e^{m_0L_0},
$$
then $\forall\, E\in[E_0 - \eta, E_0+\eta]$
$$
\|G_{\BLam_{L_0}(\uu)}(E) \| \leq e^{-m_0L_0}.
$$
Next, with $c_0 = c_0(d, \eta, m_0, L_0):= 4d+2\eta+e^{m_0L_0}$, observe that
$$\begin{array}{l}
\pr{\exists\, \ux\in \BLam_{L_0}(\uu):\, \left|W(\ux) - E_0 \right| \leq c_0}\\
= \pr{\exists\, \ux\in \BLam_{L_0}(\uu):\, \left|g[V(x_1;\om)
+ V(x_2;\om)] -[E_0 - U(\ux)] \right| \leq c_0}\\
\leq \left| \BLam_{L_0}(\uu)\right|\;
\diy{\operatornamewithlimits{\max}\limits_{\ux\in\BLam_{L_0}(\uu)}}\;
\pr{\left|V(x_1;\om) + V(x_2;\om) -g^{-1}[E_0 - U(\ux)] \right|
\leq c_0 g^{-1}}.\end{array}
$$
For $\ux=(x_1,x_2)$ with $x_1 \neq x_2$, random variables
$V(x_1;\cdot)$ and $V(x_2;\cdot)$
are independent and have a common bounded PDF ${\rm p}_V$ of compact support.
The sum $V(x_1;\cdot)+V(x_2;\cdot)$ has a bounded PDF ${\rm p}_V*{\rm p}_V$,
the convolution of ${\rm p}_V$ with itself. Thus, for
$\ux=(x_1,x_2)$ with $x_1 \neq x_2$,
$$\pr{\left|V(x_1;\om) + V(x_2;\om)-g^{-1}[E_0 - U(\ux)] \right|
\leq c_0 g^{-1}}\leq c_0 \;
\big(\diy{\operatornamewithlimits{\max}\limits\;{{\rm p}_V}*{{\rm p}_V}}\big)
\; g^{-1}.$$
For $\ux = (x_1,x_1)$, we have $V(x_1;\om) + V(x_2;\om) = 2V(x_1;\om)$, so that
$$\begin{array}{l}
\pr{\left|V(x_1;\om) + V(x_2;\om)-g^{-1}[E_0 - U(\ux)] \right|
\leq c_0 g^{-1}}\\
\quad =\pr{\left|V(x_1;\om) -(2g)^{-1}(E_0 - U(\ux)) \right|
\leq c_0 \cdot (2g)^{-1}}\leq c_0 \;\big(
\diy{\operatornamewithlimits{\max}\limits\;{{\rm p}_V}}\big)
\; g^{-1}.\end{array}$$
We see that in both cases
$\pr{\left|W(\ux) - E_0 \right| \leq c_0}\to 0$ as
$g\to\infty$, uniformly in $\ux$. Property
\Szero $\,$ then follows. $\QED$
\pmn

To complete the inductive MSA step, we will prove
\pmn

\begin{Thm}\label{MSAInd} $\forall$ given $m_0>0$, there exist
$g^*_1\in (0,+\infty)$ and $L^*_1\in (0,+\infty )$ such that
the following statement holds. Suppose that $|g|\geq g^*_1$ and
$L_0\geq L^*_1$. Then, $\forall$ $k=0,1,\ldots$ and $\forall$
interval $I\subseteq\R$, property
\SSk $\,$ implies \SSkone.
\end{Thm}
\pmn

The proof of Theorem \ref{MSAInd} occupies the rest of the paper.
Before we proceed with the proof, let us repeat that the property
\DSk $\,$
(or, equivalently, \SSk),
for $\forall$ $k\geq 0$ and $\forall$ unit interval $I\subset\R$,
follows directly from Theorems \ref{MSAInd0} and \ref{MSAInd}.

\myproof{Theorem {\rm{\ref{MSAInd}}}}
To deduce property \SSkone $\,$  from \SSk, we introduce
\pmn

\begin{Def}
Consider the following subset in $\Z^d\times\Z^d$:
$$\cD_{r_0} = \{\ux=(x_1,x_2)\in\Z^d\times\Z^d:\;
\|x_1 - x_2\|\leq r_0\}.\eqno(3.9)$$
A two-particle box $\BLam_L(\uu)$ is called interactive when
$\BLam_L(\uu) \cap \cD_{r_0} \neq \emptyset$, and non-interactive
if $\BLam_L(\uu) \cap \cD_{r_0}=\emptyset$. For a non-interactive
box $\BLam_L(\uu)$, the interaction potential $U(\ux)=0$, $\forall$
$\ux\in\BLam_{L}(\uu)$. For brevity, we use the terms {\rm{I-}}box
and {\rm{NI-}}box, respectively.
\end{Def}

The procedure of deducing property \SSkone $\,$
from \SSk $\,$
is done here separately for the following three cases.
\psn
(I) Both $\BLam_{L_{k+1}}(\ux)$ and $\BLam_{L_{k+1}}(\uy)$ are
NI-boxes.
\psn
(II) Both $\BLam_{L_{k+1}}(\ux)$ and $\BLam_{L_{k+1}}(\uy)$ are I-boxes.
\psn
(III) One of the boxes is I, while the other is NI.
\psn

In the remaining part of this section we consider case (I).
Cases (II) and  (III) are
treated in Sections 4 and  5, respectively. We repeat
that all cases require the use of property {\bf (W1)}
and/or {\bf (W2)}.

The plan for the rest of Section 3 is as follows.
We aim to derive property \SSkone $\,$
for a pair of non-interactive $L_{k+1}$-D boxes
$\BLam_{L_{k+1}}(\ux)$, $\BLam_{L_{k+1}}(\uy)$, and we
are allowed to assume
property \SSk $\,$
for every pair of $L_k$-D boxes
$\BLam_{L_k}(\widetilde{\ux})$,
$\BLam_{L_k}(\widetilde{\uy})$, where
$\ux,\uy,\widetilde{\ux}, \widetilde{\uy}\in \Z^d\times\Z^d$.
In fact, we are able to establish
property \SSkone $\,$
for non-interactive $L_{k+1}$-D
boxes $\BLam_{L_{k+1}}(\ux)$, $\BLam_{L_{k+1}}(\uy)$ directly,
without referring to \SSk.
(In cases (II) and (III)
such a reference is needed.) An important part of
our argument is a single-particle
result stated as Theorem \ref{our_Thmnt}.

Let $\BLam_{L_{k+1}}(\uu)$ be an NI-box, where
$\uu=(u_1,u_2)$. We represent it
as the Cartesian product
$$
\BLam_{L_{k+1}}(\uu) = \Lamo{L_{k+1}}(u_1) \times \Lamo{L_{k+1}}(u_2).
\eqno(3.10)
$$
Here and below, for given $\ell >1$ and
$v= (\rv^{(1)}, \ldots, \rv^{(d)})\in \R^d$:
$$
\Lamo{\ell}(v) :=
\left(\truc{\times }{d}{i=1}\left[\rv_j^{(i)}-\ell, \rv_j^{(i)}+\ell
\right]\right)\cap \Z^d.
\eqno(3.11)
$$
We call sets $\Lamo{\ell}(v)$ single-particle boxes; as before,
$\left|\Lamo{\ell}(v)\right|$ denotes the cardinality of
$\Lamo{\ell}(v)$. The boundary $\pt \Lamo{\ell}(v)$ is also defined
in a similar fashion: it is formed by the points
$y\in\Lamo{\ell}(v)$ for which $\exists$ $y'\in\Z^d\setminus\Lamo{\ell}(v)$
with $\|y-y'\|\leq 1$.

Since the potential $U$ vanishes on $\BLam_{L_{k+1}}(\uu)$,
the Hamiltonian $H^{(2)}_{\BLam_{L_{k+1}}(\uu)}$ takes
the form
$$
\begin{array}{r}H^{(2)}_{\BLam_{L_{k+1}}(\uu)}\Bphi (\ux)
=\sum\limits_{\uy\in\BLam_{L_{k+1}}(\uu):\atop{\|\uy - \ux\|=1}}
\,\Bphi(\uy)+  g\diy{\sum_{j=1,2}}V(x_j;\om)\Bphi(\ux),\\
\ux=(x_1,x_2)
\in{\BLam_{L_{k+1}}(\uu)},\qquad
\end{array}
\eqno(3.12)
$$
or, algebraically,
$$
H^{(2)}_{\BLam_{L_{k+1}}(\uu)} = H^{(1)}_{1;\Lamo{L_{k+1}}(u_1)}
\otimes {\mathbf I} + {\mathbf I}
\otimes H^{(1)}_{2;\Lamo{L_{k+1}}(u_2)}.
\eqno (3.13)
$$
Here $H^{(1)}_{j;\Lamo{L_{k+1}}(u_j)}$ is the single-particle Hamiltonian
acting on variable\\ $x_j\in\Lamo{L_{k+1}}(u_j)$, $j=1,2$:
$$\begin{array}{l}
\left(H^{(1)}_{j;\Lamo{L_{k+1}}(u_j)}\ffi\right)(x_j) =
\sum\limits_{y_j\in\Lamo{L_{k+1}}(u_j):\atop{\|y_j-x_j\|=1}}\,
\ffi(y_j)+ gV(x_j;\om)\ffi(x_j),
\end{array}\eqno(3.14)
$$
and ${\mathbf I}$ is the identity operator on the complementary variable.

Let $\psi_{j;s}(x)$ be the eigenvectors of operators
$H^{(1)}_{j;\Lamo{L_{k+1}(u_j)}}$ and $E_{j;s}$ be their
eigenvalues,
$s=1,\ldots |\Lamo{L_{k+1}}(u_j)|$. Then
the eigenvectors $\Psi_{s_1,s_2}$ of
$H^{(2)}_{\BLam_{L_{k+1}}(\uu)}$ can be represented as tensor
products:
$$
\Psi_{s_1,s_2}(\ux) = \psi_{1;s_1}(x_1)\psi_{2;s_2}(x_2),
$$
while the eigenvalues $E_{s_1,s_2}$ of
$H^{(2)}_{\BLam_{L_{k+1}}(\uu)}$ are written as sums:
$$
E_{s_1,s_2} =  E_{1;s_1} + E_{2;s_2},
$$
with $s_1= 1, \ldots, |\Lamo{L_{k+1}}(u_1)|$,
$s_2= 1, \ldots, |\Lamo{L_{k+1}}(u_2)|$.

We make use of the following definition:

\begin{Def} Fix ${\wh m}>0$ and a positive integer $\ell$.
Given $v\in\Z^d$, consider the
single-particle Hamiltonian $H^{(1)}_{\Lamo{\ell}(v)}$
in ${\Lamo{\ell}(v)}$ acting on vectors $\ffi\in\C^{\Lamo{\ell}(v)}$:
$$\begin{array}{l}
\left(H^{(1)}_{\Lamo{\ell}(v)}\ffi\right)(x) =
\sum\limits_{y\in\Lamo{\ell}(u):\atop{\|y-x\|=1}}
\ffi(y)+ gV(x;\om)\ffi(x),\;\;x\in\Lamo{\ell}(u).
\end{array}\eqno(3.15)
$$
Let $\psi_s(x)$ be the normalised eigenvectors and $E_s$
the corresponding eigenvalues of $H^{(1)}_{\Lamo{\ell}(v)}$.
We say that a single-particle box
$\Lamo{\ell}(v)$ is ${\wh m}$-non-tunnelling
(${\wh m}${\rm{-NT}}, for short), if
$$
\max_{y\in \pt \Lamo{\ell}(v)}\max\left\{
\left|  \psi_s(v)\psi_s(y) \right|:\;
E_s\in{\rm{spec}}\left(H^{(1)}_{\Lamo{\ell}(v)}\right)
\right\}  \leq e^{-{\wh m}\ell}.
\eqno (3.16)
$$
Otherwise we call it ${\wh m}$-tunnelling (${\wh m}${\rm{-T} }). A two-particle box  $\BLam_\ell(\uv)$ is called ${\wh m}$-non-tunnelling
if both of its projections $\Pi_1 \BLam_\ell(\uv)$ and $\Pi_2 \BLam_\ell(\uv)$ are ${\wh m}$-non-tunnelling.
\end{Def}

In future, the eigenvectors of finite-volume Hamiltonians
appearing in arguments and calculations, will be assumed normalised.

\psn
{\bf Remark.} Observe that (i) property ${\wh m}$-NT implies ${\wh m}'$-NT for
any ${\wh m}'\in[0,{\wh m}]$. Next, (ii)
properties ${\wh m}$-T and ${\wh m}$-NT refer only to single-particle Hamiltonians.
As we will see later, in our two-particle MSA inductive procedure,
we can use the $(2m_0)$-NT property while working with boxes
$\BLam_{L_k}(\ux)$, $\forall$ $k\geq 0$.
\pmn

The following statement gives a formal description of a property
of NI two-particle boxes
which will be referred to as property {\rm\bf(NIRoNS)}
(`non-interactive boxes are resonant or
non-singular'). As we said earlier,  property {\rm\bf(NIRoNS)}
is established for all  $k\geq 0$, by combining known results
from the single-particle localisation theory, established via
MSA or the FMM. It is worth mentioning that a property close to
{\rm\bf(NIRoNS)} was formulated in \cite{FMSS}, Proposition in
Section 6, p.43. However, the context here is different.

\begin{Lem}\label{NTLoc} Consider a pair
of single-particle boxes
$\Lamo{L_k}(u_j)$, $j=1,2,$ where $\|u_1-u_2\|>L_k +r_0$.
Given ${\wh m}>0$, assume that
$\Lamo{L_{k+1}}(u_1)$ and $\Lamo{L_{k+1}}(u_2)$ are ${\wh m}${\rm-NT}.
Next, assume that
the two-particle non-interactive box
$\BLam_{L_k}(\uu) = \Lamo{L_k}(u_1) \times \Lamo{L_k}(u_2)$
is $E${\rm-NR}. If
$L_k^{-1}\left(L_k^{\beta}  + \ln \left( (2L_k+1)^{2d}\right)\right)< 1$, then
$\BLam_{L_k}(\uu)$ is ${\wh m}^{(1)}${\rm-NS} with
$${\wh m}^{(1)} = {\wh m}\left( 1 - L_k^{-1+\beta}
- L_k^{-1}\ln \left(2L_k +1\right)^{2d}  \right).
\eqno(3.17)$$

In particular, if $ L_k^{-1}\ln \left(2L_k +1\right)^{2d}
\leq L_k^{-1+\beta}$, then
${\wh m}^{(1)}\geq {\wh m}(1- 2L_k^{-1+\beta})$.
\end{Lem}

\myproof{Lemma {\rm{3.2}}} By definition of the GFs,
$$G^{(2)}_{\BLam_{L_k}(\uu)}(\uu, \uy;E) =
\sum_{s_1=1}^{|\Lamo{L_k}(u_1)|} \sum_{s_2=1}^{|\Lamo{L_k}(u_2)|}
\frac{ \psi_{1;s_1}(u_1) \bar \psi_{1;s_1}(y_1)
\psi_{2;s_2}(u_2) \bar \psi_{2;s_2}(y_2)}
{E- \left(E_{1;s_1} + E_{2;s_2}\right)}.\eqno(3.18)$$
Here, as before, $E_{j;s}$  and $\psi_{j;s}$,
$s=1, \ldots,|\Lamo{L_k}(u_j)|$, $j=1,2$, are the eigenvalues
and the corresponding eigenvectors of
$H^{(1)}_{j;\Lamo{L_k}(u_j)}$.

Since $\Lam_{L_k}(\uu)$ is $E$-NR, the absolute values
$|E-(E_{1;s_1}+ E_{2;s_2})|$ of the denominators
in (3.18) are bounded from below by
$e^{-L_k^{\beta}}$. The sum of numerators can be bounded as follows.
First, note that if $\|\uu - \uy\|=L_k$, then either $\|u_1 - y_1\|=L_k$,
or $\|u_2 - y_2\|=L_k$. Without loss of generality, suppose
that $\|u_2 - y_2\|=L_k$, then
$$\begin{array}{l}
\left|\diy{\sum_{s_1,s_2}}
\psi_{1;s_1}(u_1) \bar \psi_{1;s_1}(y_1)
\psi_{2;s_2}(u_2) \bar \psi_{2;s_2}(y_2)
\right|\\
\qquad\qquad\leq
{\diy\sum_{s_1}}\left|\psi_{1;s_1}(u_1) \bar \psi_{1;s_1}(y_1) \right|\;
\sum_{s_2} \left|\psi_{2;s_2}(u_2)  \bar \psi_{2;s_2}(y_2) \right|\\ \;\\
\qquad\qquad\leq
|\Lamo{L_k}(u_1)|\cdot 1 \cdot |\Lamo{L_k}(u_2)| \, e^{-{\wh m}\ell}
= e^{-{\wh m}L_k} (2L_k+1)^{2d},\end{array}$$
owing to the hypothesis of non-tunnelling. Finally, we obtain
$$\left|G^{(2)}_{\BLam_{L_k}(\uu)}(\uu, \uy;E)\right|\leq
(2L_k+1)^{2d} e^{L_k^{\beta} - {\wh m}L_k} \leq e^{-{\wh m}^{(1)} L_k}.$$
This yields Lemma 3.2. $\QED$

Now introduce the following property of
single-particle Hamiltonians $H^{(1)}_{\Lamo{\ell}(v)}$:
$${\bf (NT.\,{\mbox{\boldmath${k,s}$}})}
\begin{array}{l}\quad
\pr{\hbox{ single-particle box }\;\Lamo{L_{k}}
(v)\;{\rm{is}}\;(2m_0){\rm{-NT}} }
\geq 1 - L_{k}^{-s},\end{array}\eqno(3.19)$$
where $s>0$.

Lemma \ref{NTLoc} implies the following

\begin{Lem}\label{TwoOffDiagSing} Assume property {\bf (W2)}.
Suppose that
$\forall$ $k\geq 0$, the single-particle Hamiltonians
$H^{(1)}_{\Lamo{L_{k}} (v)}$
satisfy {\bf (NT.\,{\mbox{\boldmath${k,s}$}})} with $s\geq q$:
$$\pr{ \Lamo{L_{k}} (v)\;{\rm{is}}\;(2m_0){\rm{-NT}} }
\geq 1 - L_{k}^{-q}.\eqno(3.20)$$
Suppose also that
$$L_0^{-1}\ln \left(2L_0 +1\right)^{2d} \leq
L_0^{-1+\beta} \leq \frac{1}{4}.$$
Then, $\forall$ interval $I\subseteq\R$, $\forall$
$k\geq 0$ and $\forall$ pair of non-interactive
$L_{k}${\rm -D} two-particle boxes $\BLam_{L_{k}}(\ux)$ and
$\BLam_{L_{k}}(\uy)$,
$$\pr{\exists\;E\in I:\, \BLam_{L_{k}}(\ux)
\text{ \rm and } \BLam_{L_{k}}(\uy)
\text{ \rm are } (E,m_{k}){\rm -S} } \leq 5L_{k}^{-q}.\eqno (3.21)$$
\end{Lem}
\pmn
\myproof{Lemma {\rm{\ref{TwoOffDiagSing}}}} By virtue of
Lemma 3.1,
$$\begin{array}{l}
\pr{\exists\;E\in I:\, \BLam_{L_{k}}(\ux)
\text{ \rm and } \BLam_{L_{k}}(\uy)
\text{ \rm are } (E,m_{k}){\rm -S} }\\
 \leq
\pr{\BLam_{L_{k}}(\ux) \text{ is } 2m_k{\rm -T}}
+\pr{\BLam_{L_{k}}(\uy) \text{ is } 2m_k{\rm -T}}\\
\quad + \pr{\exists\;E\in I:\, \BLam_{L_{k}}(\ux)
\text{ \rm and } \BLam_{L_{k}}(\uy)
\text{ \rm are } E{\rm -R} }\\
\leq 2\cdot 2 L_k^{-s} + L_k^{-q} \leq 5 L_k^{-q}. \qquad\QED
\end{array}$$
\pmn

The validity of (3.20) is guaranteed by
\pmn

\begin{Thm}\label{our_Thmnt}
Consider single-particle Hamiltonians $H^{(1)}_{\Lamo{L_k}(v)}$,
$v\in\Z^d$, $k=0, 1, \ldots$. Then $\exists$ $g^*_2,
L^*_2\in (0,+\infty)$ such that when
$|g|\geq g^*_2$ and $L_0\geq L^*_2$,
the following bound holds true for all $k\geq 0$:
$$
\pr{ \Lamo{L_{k}} (v)\;{\rm{is}}\; (2m_0){\rm{-NT}} }
\geq 1 - L_{k}^{-s},
\; s = \frac{\ptt - 2(1+\alpha)d}{\alpha}.\eqno(3.22)$$
\end{Thm}
\pmn

In other words, Theorem \ref{our_Thmnt} asserts property
{\bf (NT.\,{\mbox{\boldmath${k,s}$}})} with
$s=$\\ $\big[\ptt - 2(1+\alpha)d\big]\big/\alpha$. So, it
suffices to assume that
$\ptt \geq \alpha q + 2(1+\alpha)d$. Since, as we observed before,
$\ptt = \ptt(g) \to\infty$ as $|g|\to\infty$, the latter inequality
holds for $|g|$ large enough.

As was said in Section 1, the reader can find, in the
forthcoming manuscript \cite{CS2},
a proof of Theorem \ref{our_Thmnt} based on an adaptation of MSA techniques
from \cite{DK} and valid under the IID assumption and condition (1.3).
However, a stronger estimate was proved
in \cite{A94}, with the help of the FMM, under condition (1.2) (in fact, the
assumptions on the
external potential $V(x;\om )$, $x\in\Z^d$, adopted in
\cite{A94} are more general than IID, and, according to
\cite{A08}, they can be further relaxed; see \cite{AW}). Namely, bound (1.6) from \cite{A94}
implies that
$$\pr{ \Lamo{L_{k}} (v)\;{\rm{is}}\; (2m_0){\rm{-NT}} }
\geq 1 - e^{-{\wt m} L_{k}},\eqno(3.23)$$
where ${\wt m}={\wt m}(g)\to \infty$ as $|g|\to\infty$.
We also recall that, for one-dimensional single-particle
models, exponential bounds of probability of exponential
decay of eigen-functions
in finite volumes were obtained in \cite{GMP} (for Schr\"{o}dinger
operators on $\R$) and in \cite{KS} (for lattice Schr\"{o}dinger
operators on $\Z$).

We thus come to the following conclusion.
\pmn

\begin{Thm} $\forall$ given interval $I\subseteq\R$
and $k=0,1,\ldots$, property \SSk $\,$
holds for all pairs of $L_k${\rm -D} non-interactive boxes
$\BLam_{L_k}(\ux)$, $\BLam_{L_k}(\uy)$.
\end{Thm}

Summarising the above argument: the validity of property \SSkone $\,$
for a pair of two-particle
NI-boxes did not require us to assume \SSk.
However, in the course of deriving \SSkone $\,$
for NI-boxes we used
property (3.20) for single-particle boxes, as well as the Wegner-type
property {\bf (W2)}.

This completes the analysis of the case (I) where both boxes
$\BLam_{L_{k+1}}(\ux)$ and $\BLam_{L_{k+1}}(\uy)$ are NI.

For future use, we also give

\begin{Lem}\label{LemonM}
Consider a two-particle box $\BLam_{L_{k+1}}(\uu)$.
Let $M(\BLam_{L_{k+1}}(\uu);E)$ be the maximal number of $(E,m_k)${\rm -S},
pair-wise $L_k${\rm{-D}}$\;$
{\rm{NI-}}boxes $\BLam_{L_k}(\uu^{(j)})\subset \Lam_{L_{k+1}}(\uu)$.
The following property holds
$$
\pr{\exists E\in I:\;\; M(\BLam_{L_{k+1}}(\uu);E) \geq 2 }
\leq  L_k^{2d(1+\alpha)} \cdot  5 L_k^{-2\ptt}
<   L_k^{4d\alpha} \cdot  L_k^{-2\ptt}.\eqno (3.24)$$
\end{Lem}

\myproof{Lemma {\rm{\ref{LemonM}}}} The number of possible pairs
of centres $\uu^{(1)}, \uu^{(2)}$ is bounded by
$$(2L_{k+1}+1)^{2d}  \leq (2L_k^{\alpha}+1)^{2d}
\leq (L_k^{\alpha+1})^{2d} = L_k^{2d(\alpha+1)},$$
while for a given pair of centres one can apply Theorem
\ref{our_Thmnt}. $\QED$

\psn

\section{Interactive pairs of singular boxes}
\label{Case_II}

Speaking informally, case (II) corresponds to a two-particle system
with `confinement': in both boxes $\BLam_{L_{k+1}}(\ux)$ and
$\BLam_{L_{k+1}}(\uy)$, particles are at a
distance $\leq 2L_k + r_0$ from each other, and form a `compound quantum
object' which
can be considered as a `single particle' subject to a random
external potential. It is not entirely surprising, then,
that such a compound object should feature localisation properties
resembling those from the single-particle theory. The reader may see that
the analysis needed to cover case (II) is rather similar to that
in [DK]. It relies essentially upon properties {\bf (W1)} and {\bf (W2)}
(Wegner-type estimates).
However, it is worth mentioning that the derivation of estimates
{\bf (W1)} and {\bf (W2)} required new ideas due to strong
dependencies in the random
potential $gV(x_1;\om) + gV(x_2;\om)$. As was said before, these
dependencies do not
decay as $\|x_1-x_2\|\to\infty$. Our proofs given in \cite{CS1}
are based on Stollmann's lemma (cf. \cite{St1}, \cite{St2})
rather than on the original ideas of Wegner.

The main outcome in case (II) is Theorem \ref{ThmTwoISing} placed at
the end of this section. Before we proceed further, let us state
a geometric assertion (see Lemma \ref{DistDiag} below)
which we prove in Section 6. Given a two-particle box $\BLam_L(\uu )$,
with $\uu=(u_1, u_2)$, and $u_j=(\ru^{(1)}_j, \ldots,
\ru^{(d)}_j)\in\Z^d$, set
$$
\Pi \, \BLam_{L}(\uu ) = \Pi_1 \, \BLam_{L}(\uu )
\cup \Pi_2 \, \BLam_{L}(\uu )
\subset \Z^d.
\eqno (4.1)
$$
Here $\Pi_1 \, \BLam_{L}(\uu )$ and $\Pi_2 \, \BLam_{L}(\uu )$
denote the projections of $\BLam_L(\uu )$ to the first and the second
factor in $\Z^d\times\Z^d$:
$$
\Pi_j\,\BLam_L(\uu) = \left(
\truc{\times }{d}{i=1}
[\ru_j^{(i)}-L, \ru_j^{(i)}+L ]\right)\cap\Z^d,\;\;j=1,2;
$$
cf. (1.9).
In other words, $\Pi_j\,\BLam_L(\uu)$ describes a `supporting domain'
of the single-particle external potential $\{V(x),\;x\in\Z^d\}$
contributing into the potential field $W(\ux )$, $\ux\in\BLam_L(\uu)$.
\pmn

\begin{Lem}\label{DistDiag}
Let be $L> r_0$ and consider two interactive $8L$-D
boxes $\BLam_L(\uu')$ and $\BLam_L(\uu'')$, with
$\dist\left[\BLam_L(\uu'),\BLam_L(\uu'')\right] >8L$. Then
$$\Pi\BLam_L(\uu') \cap \Pi\BLam_L(\uu'')= \emptyset.\eqno (4.2)$$
\end{Lem}
\pmn

Lemma \ref{DistDiag} is used in the proof of Lemma \ref{ProbISing}
which, in turn, is important in establishing
Theorem \ref{ThmTwoISing}. Actually, it is a natural complement to
Lemma 2.2 in \cite{CS1}.
Let $I\subseteq\R$ be an interval. Consider the following assertion
$$\ISk:
\begin{array}{l}\hbox{$\forall$ pair of  interactive $L_k${\rm{-D}} boxes
$\BLam_{L_k}(\ux )$ and $\BLam_{L_k}(\uy )$:}\\
\P\;\Big\{ \exists \, E\in I:\,{\rm{both}}\;
\BLam_{L_k}(\ux ),\;\BLam_{L_k}(\uy )\;{\rm{are}}
\;(E,m_k)\text{-S}\Big\}\leq L_k^{-2p}.\end{array}
\eqno (4.3)$$
\pmn

\begin{Lem}\label{ProbISing}
Given $k\geq 0$, assume that property
{\bf (IS.{\mbox{\boldmath${k}$}})} holds true. Consider a box
$\BLam_{L_{k+1}}(\uu )$
and let $N(\BLam_{L_{k+1}}(\uu );E)$ be the maximal number of
$(E,m_k)${\rm -S},
pair-wise $L_k${\rm{-D}}\; {\rm{I-}}boxes
$\BLam_{L_k}(\uu^{(j)})\subset\BLam_{L_{k+1}}(\uu )$.
Then $\forall$ $n\geq 1$,
$$
\pr{\exists\;E\in I:\;\; N(\BLam_{L_{k+1}}(\uu );E) \geq 2n }\leq
L_k^{2n(1+d\alpha)} \cdot  L_k^{-2np}.
\eqno (4.4)
$$
\end{Lem}
\pmn

\myproof{Lemma {\rm{\ref{ProbISing}}}}Suppose
$\exists$ I-boxes $\BLam_{L_k}(\uu^{(1)}),\ldots$,
$\BLam_{L_k}(\uu^{(2n)})$\\ $\subset\BLam_{L_{k+1}}(\uu )$ such that
any two of them are $L_k$-D, i.e., are at the distance
$>8L_k$.

By virtue of Lemma \ref{DistDiag}, it is readily seen that

(a) $\forall$ pair $\BLam_{L_k}(u^{(2i-1)})$, $\BLam_{L_k}(u^{(2i)})$,
the respective (random) operators\\
$H^{(2)}_{\BLam_{L_k}(\uu^{(2i-1)})}(\om )$
and $H^{(2)}_{\BLam_{L_k}(\uu^{(2i)})}(\om )$ are independent,
and so are their
spectra and GFs.

(b) Moreover, the pairs of operators,
$$\left(H^{(2)}_{\BLam_{L_k}(\uu^{(2i-1)})}(\om ),
H^{(2)}_{\BLam_{L_k}(\uu^{(2i)})}(\om )\right),\;\;i=1, \dots, n,\eqno(4.5)$$
form an independent family.

Indeed, operator
$H^{(2)}_{\BLam_{L_k}(\uu^{(i)})}$, with $i\in\{1,\ldots, 2n\}$, is
measurable relative to the sigma-algebra
$\ccB_i$  generated by random variables
$\{V(x), \, x\in \Pi \, \BLam_{L_k}(\uu^{(i)})\}$, with
$$
\Pi \, \BLam_{L_k}(\uu^{(i)}) = \Pi_1 \, \BLam_{L_k}(\uu^{(i)})
\cup \Pi_2 \, \BLam_{L_k}(\uu^{(i)})\subset \Z^d.$$
Now, by Lemma 4.2, the sets $\Pi \, \BLam_{L_k}(\uu^{(i)})$,
$i\in\{1,\ldots , 2n\}$, are pairwise disjoint,
so that all sigma-algebras $\ccB_i$, $i\in\{1,\ldots ,2n\}$, are
independent.
\psn

{\bf Remark.} This property formalises the observation made
in the beginning of this section: a pair of particles
corresponding to an interactive box of size $2L_k$ forms a
"compound quantum object" of size $< 8L_k$, and their
analysis is quite similar to that from the single-particle MSA.
\psn

Thus, any collection of events $\cA_1$,
$\ldots$, $\cA_{n-1}$ related to the corresponding pairs
$\left(H^{(2)}_{\BLam_{L_k}(\uu^{(2i-1)})},H^{(2)}_{\BLam_{L_k}(\uu^{(2i)})}
\right)$, $i=1, \ldots,$ $n$, also form an independent family.

Now, for $i=1,\ldots, n-1$,  set
$$
\cA_i = \myset{ \exists \, E\in I:\,\hbox{both }
\BLam_{L_k}(\uu^{(2i-1)}),\;\;\BLam_{L_k}(\uu^{(2j+2)})\;{\rm{are}}\;
(E,m_k)\text{-S} }.
\eqno (4.6)
$$
Then, by virtue of \ISk,
$$
\P\;\Big\{ \cA_j\Big\}\leq L_k^{-2p},\;\;0\leq j\leq n-1,
\eqno (4.7)
$$
and by virtue of independence of events $\cA_0$,
$\ldots$, $\cA_{n-1}$, we obtain
$$
\P\;\Big\{ \bigcap_{j=0}^{n-1} \cA_j\Big\} =
\prod_{j=0}^{n-1}\P\;\Big\{\cA_j\Big\}\leq
\left(L_k^{-2p}\right)^{n}.
\eqno (4.8)
$$
To complete the proof, note that the
total number of different families
of $2n$ boxes $\BLam_{L_k}\subset\BLam_{L_{k+1}}(\uu )$
with required properties is bounded from above by
$$
\frac{1}{(2n)!} \left( 2 (L_k + r_0 + 1) L_{k+1}^{d}\right)^{2n}
\leq \frac{1}{(2n)!} \left(4 L_k L_{k+1}^{d} \right)^{2n}
\leq L_k^{2n(1+d\alpha)},
$$
since their centres must belong to the subset
$\cD_{L_k+r_0}\cap \BLam_{L_{k+1}}(\uu )$.
Here
$$\cD_{L_k+r_0}=\myset{(x_1,x_2)\in\Z^d\times\Z^d:\,\|x_1 - x_2\|
\leq L_k+r_0}
$$
is a `layer' of width $2(L_k+r_0)$ adjoint to the set $\cD := \{\ux =(x,x),\; x\in \Z^d\}$,
the diagonal in $\Z^d\times\Z^d$.

Recall also that $r_0 < L_0\leq L_k$,
$k\geq 0$, by our assumption. This yields Lemma \ref{ProbISing}.
$\qquad\QED$

\begin{Lem}\label{HowManySub} Let $K(\BLam_{L_{k+1}(\uu )};E)$ be the maximal
number of $(E,m_k)${\rm -S}, pair-wise $L_k${\rm{-D}}
boxes $\BLam_{L_k}(\uu^{(j)})\subset \BLam_{L_{k+1}}(\uu )$ (interactive
or non-interactive).
Then $\forall$ $n\geq 1$,
$$
\pr{\exists E\in I:\;\; K(\BLam_{L_{k+1}(\uu )};E) \geq 2n +2}\leq
L_k^{4d\alpha} \cdot  L_k^{-2\ptt} +
L_k^{2n(1+d\alpha)} \cdot  L_k^{-2np}.
\eqno (4.9)$$
\end{Lem}
\myproof{Lemma {\rm{\ref{HowManySub}}}}
Assume that $K(\BLam_{L_{k+1}(\uu )};E) \geq 2n+2$.
Let $M(\BLam_{L_{k+1}(\uu )};E)$ be as in Lemma \ref{LemonM}
and $N(\BLam_{L_{k+1}(\uu )};E)$ as in Lemma \ref{ProbISing}. Obviously,
$$K(\BLam_{L_{k+1}(\uu )};E) \leq M(\BLam_{L_{k+1}(\uu )};E)
+ N(\BLam_{L_{k+1}(\uu )};E).$$
Then either $M(\BLam_{L_{k+1}(\uu )};E)\geq 2$ or
$N(\BLam_{L_{k+1}(\uu )};E)\geq 2n$. Therefore,
$$\begin{array}{l}
\pr{\exists E\in I:\;\; K(\BLam_{L_{k+1}(\uu )};E) \geq 2n +2}\\
\qquad\leq \pr{\exists E\in I:\;\; M(\BLam_{L_{k+1}(\uu )};E) \geq 2}\\
\qquad\;\;+\;
\pr{\exists E\in I:\;\; N(\BLam_{L_{k+1}(\uu )};E) \geq 2n}\\
\qquad\leq L_k^{4d\alpha} \cdot  L_k^{-2\ptt} +
L_k^{2n(1+d\alpha)} \cdot  L_k^{-2np},
\end{array}
$$
by virtue of (3.24) and (4.4)
$\QED$

An elementary calculation now gives rise to the following
\pmn

\begin{Cor}\label{CorHowManySub} Under assumptions of Lemma
{\rm{\ref{HowManySub}}}, with $n\geq 4$, $p\geq 12d+9$,
$\ptt \geq 3p+3d$, $\alpha = \valpha$, for $L_0 \geq 2$ large enough,
we have
$$
\pr{\exists E\in I:\;\; K(\BLam_{L_{k+1}(\uu )};E) \geq 2n +2}\leq
\,   L_{k+1}^{-2p-1}.\eqno (4.10)$$
\end{Cor}
\pmn

{\bf Remark.} Our lower bounds on values of $n$, $p$ and $\ptt$ are {\it not}
sharp.
\psn
{\bf Definition 4.1.}\label{CNR}
A box $\BLam_{L_{k+1}}(\uv)$ is called $(E,J)$-completely non-resonant
($(E,J)$-CNR in brief), if
the following properties are fulfilled:
\psn
{\rm (i)} $\BLam_{L_{k+1}}(\uv)$ is $E$-NR;
\psn
{\rm (ii)} all boxes of the form
$\BLam_{j(L_k+1)}(\uy)\subset \BLam_{L_{k+1}}(\uv)$,
$\uy\in\BLam_{L_{k+1}}(\uv)$, $j=1, \ldots, J$, are
$E$-NR.
\pmn

As follows from Definition \ref{CNR} and property
{\bf (W.2)}, we have
\pmn

\begin{Lem}\label{LemCNR} Let $
= \BLam_{L_{k+1}}(\uu')$, $
= \BLam_{L_{k+1}}(\uu'')$ be two $L_{k+1}${\rm -D} boxes.
Then, for $L_0> (J+1)^2$,
$$\begin{array}{r}
\pr{\forall\, E\in I:\, \text{ either } \BLam_{L_{k+1}}(\uu')
 \text { or } \BLam_{L_{k+1}}(\uu'') \text{ is } (E,J){\rm-CNR}}
\qquad\\
\geq 1 - (J+1)^2L_{k+1}^{-(q\alpha^{-1}-2\alpha)}
> 1 - L_{k+1}^{-(q'-4)}, \; q':= q/\alpha.\\
\end{array}
\eqno (4.11)
$$
\end{Lem}
\pmn

The statement of Lemma \ref{J_NS} below
is a simple reformulation of Lemma 4.2
from \cite{DK}, adapted to our notations.
Indeed, the reader familiar with
the proof given in \cite{DK} can see that the structure
of the external potential is irrelevant to
this completely deterministic statement. So it applies
directly to our model with potential
$U(\ux) + gW(\ux )$. For that reason, the proof of
 Lemma \ref{J_NS} is omitted.

\begin{Lem}\label{J_NS}  Fix an odd positive integer $J$
and suppose that the following properties are fulfilled:
\psn
\centerline{{\rm (i)} $\BLam_{L_{k+1}}(\uv)$ is $(E,J)${\rm{-CNR}}, and
{\rm (ii)} $K(\BLam_{L_{k+1}(\uu )};E)\leq J$.}
\psn

Then for sufficiently large $L_0$, box $\BLam_{L_{k+1}}(\uv)$
is $(E,m_{k+1})${\rm{-NS}} with
$$m_{k+1} \geq m_k \left( 1 - \frac{5J+6}{(2L_k)^{1/2}} \right)
 > m_0/2>0.\eqno (4.12)$$
\end{Lem}

{\bf Remark.} In \cite{DK}, it is also assumed that
$\alpha < (J+1)(d + \hhalf)$.
In our case, this is automatically satisfied
with $\alpha=3/2$ and $J\geq 1$.
In particular, with $J=9$, we obtain
$$
m_{k+1} \geq m_k \left( 1 - \frac{51}{(2L_k)^{1/2}} \right) >
m_k \left( 1 - \frac{40}{L_k^{1/2}} \right),
\eqno{(4.13)}
$$
which explains our assumption (3.1) and the recursive
definition (1.13)
with $\gamma=40$.
\psn
Now comes a statement which extends Lemma 4.1 from \cite{DK} to pairs
of two-particle $\;L_k$-D$\;$
I-boxes.
\pmn
\begin{Thm}\label{ThmTwoISing}
$\forall$ given interval $I\subseteq\R$, there exists
$L^*_3\in(0,+\infty)$ such that if
$L_0\geq L^*_0$, then, $\forall\,
k\geq 0$, property \ISk  $\,$
in {\rm{(4.1)}}
implies \ISkone  $\,$.
\end{Thm}
\pmn
{\it Proof of Theorem} \ref{ThmTwoISing}. Let $\ux,\uy\in\Z^d\times\Z^d$
and assume that
$\BLam_{L_{k+1}}(\ux )$ and $\BLam_{L_{k+1}}(\uy )$ are
$L_k${\rm{-D}} I-boxes. Consider the following two events:
$$\cB = \Big\{ \exists \, E\in I:\,{\rm{both}}\;
\BLam_{L_{k+1}}(\ux )
\text{ and } \BLam_{L_{k+1}}(\uy )\;
\text{ are } (E,m_{k+1})\text{-S}\Big\}\,,$$
and, for a given odd integer $J$,
$$\cR = \Big\{\exists \, E\in I:\, \text{ neither }
\BLam_{L_{k+1}}(\ux ) \text{ nor } \BLam_{L_{k+1}}(\uy )
\text{ is } (E,J){\rm{-CNR}} \Big\}.$$
By virtue of Lemma  \ref{LemCNR}, we have,
with $L_0$ large enough ($L_0\geq J+1)^2$) and $\alpha = 3/2$:
$$\pr{\cR} < L_{k+1}^{-(q'-4)}, \; q':= q/\alpha.\eqno(4.14)$$

Further,
$$\pr{\cB} = \pr{ \cB\cap \cR} + \pr{\cB\cap\cR^{\rm c}}
\leq \pr{\cR} + \pr{\cB\cap\cR^{\rm c}},$$
and we know that $\pr{\cR}\leq L_{k+1}^{-q'+4}$.
So, it suffices now to estimate $\pr{\cB\cap \cR^{\rm c}}$. Within
the event
$\cB\cap \cR^{\rm c}$, for any $E\in I$, one of
the boxes $\BLam_{L_{k+1}}(\ux )$, $\BLam_{L_{k+1}}(\uy )$ must be $(E,J)$
-CNR.
Without loss of generality,
assume that for some $E\in I$, $\BLam_{L_{k+1}}(\ux )$ is
$(E,J)$-CNR and $(E,m_{k+1})$-S. By Lemma
\ref{J_NS}, for such value of $E$, $K(\BLam_{L_{k+1}}(\ux );E) \geq J+1$.
We see that
$$
\cB \cap \cR^{\rm c} \subset \Big\{\exists E\in I:\;\;
K(\BLam_{L_{k+1}}(\ux );E) \geq J+1 \Big\}
$$
and, therefore, by Lemma \ref{HowManySub}, with the same
values of parameters  as in Corollary
\ref{CorHowManySub}, as before:
$$
\pr{\cB \cap \cR^{\rm c}} \leq \pr{\exists E\in I:\;\;
K(\BLam_{L_{k+1}}(\ux );E) \geq J+1}
\leq L_k^{-2p}.
\eqno (4.15)
$$
\qquad$\QED$
\pmn

In what follows we consider $J=9$ although it will be convenient
to use symbol $J$, in particular, to stress analogies with
\cite{DK}.
\pmn

\section{Mixed pairs of singular two-particle boxes}
\label{Case_III}

It remains to derive the property \SSkone $\,$
in case (III),
i.e., for mixed pairs
of two-particle boxes (where one is I and the other NI).
Here we use several properties which have
been established earlier in this paper for all scale lengths,
namely, {\bf(W1)}, {\bf(W2)}, {\bf (NT.\,{\mbox{\boldmath${k,s}$}})}
with $s\geq q$, {\bf(NIRoNS)},
and the inductive assumption \ISkone
which we have already
derived from \ISk $\,$
in Section 4.

A natural counterpart of Theorem \ref{ThmTwoISing} for
mixed pairs of boxes is the following

\begin{Thm}\label{ThmINISing} $\forall$ given interval $I\subseteq\R$,
there exists a constant
$L^*_4\in(0,+\infty)$ with the following property. Assume that
$L_0\geq L^*_4$ and, for a given $k\geq 0$, the property \SSk $\,$
holds:
{\rm (i)} $\forall$ pair of $L_k${\rm -D} $\;${\rm{NI-}}boxes
$\BLam_{L_k}({\wt\ux})$,
$\BLam_{L_k}({\wt\uy})$, and
{\rm (ii)} $\forall$ pair of $L_k${\rm -D} $\;${\rm{I-}}boxes
$\BLam_{L_k}({\wt\ux})$, $\BLam_{L_k}({\wt\uy})$.

Let $\BLam_{L_{k+1}}(\ux)$,
$\BLam_{L_{k+1}}(\uy)$ be a pair of $L_{k+1}${\rm -D} boxes, where
$\BLam_{L_{k+1}}(\ux)$ is $\;${\rm I}$\;$ and
$\BLam_{L_{k+1}}(\uy)\;$ {\rm{NI}}. Then
$$\P\;\Big\{\exists \, E\in I:\,\text{\rm both } \BLam_{L_{k+1}}(\ux),
\;\;
\BLam_{L_{k+1}}(\uy )\;{\rm{are}}\;(E,m_{k+1}){\rm -S}\Big\}
\leq L_{k+1}^{-2p}.\eqno (5.1)$$
\end{Thm}
\pmn

Before starting a formal proof
we give an informal description of our strategy.
\begin{enumerate}
\item We are going to list several situations which may give rise
to singularity of a mixed pair $\BLam_{L_{k+1}}(\ux)$ (an I-box),
$\BLam_{L_{k+1}}(\uy )$ (an NI-box).
Next, we show that each situation is covered by an event of (negligibly)
small probability. Finally, we show
that if neither of these events occurs, the pair of boxes in question
cannot be $(E,m_{k+1})$-S.

\item Given a pair of an I-box and an NI-box, which are $(E,m)$-S
for some (and the same) $E$, we note first
that, owing to {\bf (NIRoNS)}, with high probability, the NI-box has
to be $E$-R.
If it is not, we count such an event as an unlikely
situation which may give rise to simultaneous singularity
of the pair in question.

\item Assuming that the NI-box $\BLam_{L_{k+1}}(\uy )$
is $E$-R, we apply the Wegner-type estimate
{\bf (W2)} and conclude that, with high
probability, neither the I-box $\BLam_{L_{k+1}}(\ux)$,
nor any of its sub-boxes of size
$2L_k$ is $E$-R. Again, the presence of
`unwanted' $E$-R boxes is considered as an unlikely situation.
Otherwise, we conclude that $\BLam_{L_{k+1}}(\ux)$
is $(E,J)$-CNR.

\item Focusing on the I-box $\BLam_{L_{k+1}}(\ux)$,
we use properties {\bf (W2)} and \ISk  $\,$
to prove that, with high
probability, it contains a limited number of distant sub-boxes of size
$2L_k$ which are $(E,m_k)$-S. Specifically, it is unlikely
that $\BLam_{L_{k+1}}(\ux)$ contains at least two $L_k$-D$\;$
NI-sub-boxes of size $2L_k$ (by
{\bf (NIRoNS)} and {\bf (W2)}); it is also unlikely
that it contains at least $(J-1)$ $L_k$-D $\;$I-sub-boxes of
size $2L_k$, by virtue of \ISk.

\item Finally, if a two-particle box of width $2L_{k+1}$ is
both $(E,J)$-CNR
and contains at
most $(J-2) + (2-1)=J-1$ distant sub-boxes, it must be $(E,m_{k+1})$-NS,
which is a possibility outside the event in Eqn (5.1).
So, the sum of
probabilities of the above-mentioned events
gives an upper bound for the probability of simultaneous
singularity of the given mixed pair of boxes.
\end{enumerate}

\myproof{Theorem {\rm\ref{ThmINISing}}} Recall that the Hamiltonian
$H^{(2)}_{\BLam_{L_{k+1}}(\uy )}$ is decomposed as in Eqns
(3.12), (3.13). Consider the following three events:
$$\cB = \Big\{ \exists \, E\in I:\,{\rm{both}}\;
\BLam_{L_{k+1}}(\ux),\;\;\BLam_{L_{k+1}}(\uy)\;
\text{ are } (E,m_{k+1})\text{-S}\Big\}\,,$$
$$\cT = \Big\{\hbox{either $\Lam_{L_{k+1}}(y_1)$
or  $\Lam_{L_{k+1}}(y_2)$ is $(2m_0)$-T} \Big\},$$
and
$$\cR = \Big\{\exists \, E\in I:\, \text{ neither }
\BLam_{L_{k+1}}(\ux) \text{ nor }
\BLam_{L_{k+1}}(\uy)\text{ is } (E,J){\text{-CNR}} \Big\}.$$
Event $\;$B$\;$ is the one figuring in the bound (5.1), and
we are interested in estimating its probability.

Recall that by virtue of (3.22), we have
$$\pr{ \cT } \leq L_{k+1}^{-s},\;\hbox{ where }\;
s=\frac{\ptt -2(1+\alpha)d}{\alpha}=\frac{\ptt -5d}{\alpha},
\eqno (5.2)$$
while for event $\cR$ we have again, by virtue of
Lemma \ref{LemCNR} and inequality (4.13),
with our choice of parameters
$J$ and $L_0$ ($J=9$ and $L_0$ large enough),
$$\pr{\cR} \leq L_{k+1}^{-q+2}.\eqno (5.3)$$
Further,
$$\begin{array}{cl}
\pr{\cB} &= \pr{ \cB\cap\cT} + \pr{ \cB\cap \cT^{\rm c}}\\
\;&
\leq \pr{T} + \pr{\cB\cap \cT^{\rm c}} \leq
L_{k+1}^{-s}+ \pr{\cB\cap \cT^{\rm c}}.\end{array}
$$
Now, we estimate $\pr{\cB\cap \cT^{\rm c}}$:
$$\begin{array}{cl}
\pr{\cB \cap \cT^{\rm c}}&= \pr{ \cB \cap \cT^{\rm c}
\cap \cR} + \pr{\cB\cap \cT^{\rm c} \cap \cR^{\rm c}}\\
\;&
\leq \pr{\cR} + \pr{\cB\cap \cT^{\rm c}\cap \cR^{\rm c}}
\leq L_{k+1}^{-q+2}
+ \pr{\cB\cap \cT^{\rm c}\cap \cR^{\rm c}}.\end{array}$$

So, it suffices to estimate $\pr{\cB\cap \cT^{\rm c} \cap
\cR^{\rm c}}$. Within the event
$\cB\cap \cT^{\rm c} \cap \cR^{\rm c}$,
one of the boxes $\BLam_{L_{k+1}}(\ux)$, $\BLam_{L_{k+1}}(\uy)$
is $E$-NR. It cannot be the NI-box
$\BLam_{L_{k+1}}(\uy)$. Indeed, by
Corollary 4.1, had box $\BLam_{L_{k+1}}(\uy)$
been both $E$-NR and $(2m_0)$-NT,
it would have been $(E,m_{k+1})$-NS, which is not allowed
within the event $\cB$.
Thus, the I-box $\BLam_{L_{k+1}}(\ux)$ must be $E$-NR,
but $(E,m_{k+1})$-S:
$$\cB\cap \cT^{\rm c} \cap \cR^{\rm c} \subset
\{\exists \, E\in I:\;\BLam_{L_{k+1}}(\ux)
\text{ is } (E,m_{k+1}){\text{-S}} \text{ and } E{\text{-NR}} \}.$$
However, applying Lemma \ref{J_NS}, we see that
$$\begin{array}{r}
\{\exists \, E\in I:\,
\BLam_{L_{k+1}}(\ux) \text{ is }(E,m_{k+1}){\text{-S}} \text{ and }E{
\text{-NR}}\}\qquad\qquad\\
\subset \{\exists \, E\in I:\, K(\BLam_{L_{k+1}}(\ux);E) \geq J+1\}.
\end{array}$$
Therefore, with the same values of parameters
as in Corollary \ref{CorHowManySub},
$$\begin{array}{cl}
\pr{\cB\cap \cT^{\rm c}\cap \cR^{\rm c}}
&\leq \pr {\exists \, E\in I:\, K(\BLam_{L_{k+1}}(\ux)
;E) \geq 2n+2 }\\
\;&\leq 2L_{k+1}^{-1}\,L_{k+1}^{-2p}.\end{array}\eqno (5.4)$$
Finally, we get, with $q':=q/\alpha$,
$$\begin{array}{cl}
\pr{\cB} &\leq \pr{\cB\cap \cT} + \pr{\cR}  + \pr{\cB\cap \cT^{\rm c}
\cap \cR^{\rm c}}\\
\;&\leq L_{k+1}^{-s} + L_{k+1}^{-q'+4} + 2L_{k+1}^{-1} \, L_{k+1}^{-2p}
\leq L_{k+1}^{-2p},\end{array}\eqno (5.5)$$
if we can guarantee that
$$\max \left\{ L_{k+1}^{-s+2p}, L_{k+1}^{-q'+4+2p},
2L_{k+1}^{-1} \right\} \leq \frac{1}{3}.\eqno (5.6)$$
The bound (5.6) follows from our assumptions, provided that
$L_0$ is large enough and
$$s-2p=\frac{\ptt -5d}{\alpha}-2p >1,\;\;q'-2p-4 >1.\eqno (5.7)$$
This completes the proof of Theorem \ref{ThmINISing}.
$\qquad\QED$
\pmn

Therefore, Theorem \ref{MSAInd} is proven. In turn, this
completes the proof of Theorem \ref{MThm}.
\bigskip

\section{Proof of Lemma \ref{DistDiag} }

Recall that we deal with two-particle boxes $\BLam' := \BLam_L(\uu')$
and $\BLam'' := \BLam_L(\uu'')$ such that
\psn
\centerline{{\rm (i)} $\dist\left( \BLam', \BLam''\right) >8L$
and
{\rm (ii)} $\BLam'\cap \cD_{r_0} \neq \emptyset\neq
\BLam''\cap \cD_{r_0}$.}
\psn
Recall that we denote by $\cD$ the diagonal in
$\Z^d\times\Z^d$:  $\cD= \{\ux =(x,x),\; x\in \Z^d\}$.
Then property (ii) implies that
$$\BLam_{L+r_0}(\uu') \cap \cD \neq \emptyset,
\; \BLam_{L+r_0}(\uu'') \cap \cD \neq \emptyset,\eqno(6.1)$$
so that $\exists\, \widetilde \ux' = (\widetilde x', \widetilde x')\in
\BLam_{L+r_0}(\uu') $ and
$\exists\, \widetilde \ux'' = (\widetilde x'', \widetilde x'')
\in \BLam_{L+r_0}(\uu'') $. Next, observe that
$$\dist( \BLam_{L+r_0}(\uu'), \BLam_{L+r_0}(\uu''))
\geq \dist( \BLam_{L}(\uu'), \BLam_{L}(\uu'')) - 2r_0 > 8L - 2r_0 >0,
\eqno(6.2)$$
owing to the assumption $L> r_0$, and therefore,
$$\|\widetilde \ux' - \widetilde \ux''\| = \|\widetilde x'_1
- \widetilde x''_1\|
= \|\widetilde x'_2 - \widetilde x''_2\| > 8L - 2r_0,
\eqno(6.3)$$
since $\widetilde \ux', \widetilde \ux'' \in\cD$.

Further, for arbitrary points $\ux'\in\BLam'$,
$\ux''\in\BLam''$, and any $j\in\{1,2\}$,
we can write the triangle inequality as follows:
$$\dist( \widetilde x'_j, \widetilde x''_j)
\leq \dist( \widetilde x'_j, x'_j) + \dist( x'_j, x''_j)
+ \dist( x''_j, \widetilde x''_j)$$
or, equivalently,
$$\begin{array}{cl}
\dist( x'_j, x''_j)
\geq \dist( \widetilde x'_j, \widetilde x''_j) - \dist( \widetilde x'_j, x'_j) - \dist( x''_j, \widetilde x''_j) \\
> 8L - 2r_0 - (2L + 2r_0) - (2L + 2r_0) = 4L - 6r_0 \geq 2L > 0,
\end{array}\eqno(6.4)$$
since
$$
\dist( \widetilde x'_j, x'_j) \leq \diam(\BLam_{L+r_0})=2L+2r_0
$$
and the same upper bound holds for $\dist( x''_j, \widetilde x''_j)$.
We see that, for $j=1,2$,
$$\dist( \Pi_j \BLam', \Pi_j \BLam'') > 2L >0,\eqno(6.5)$$
so that $\Pi_1 \BLam' \cap \Pi_1 \BLam'' = \emptyset$,
$\Pi_2 \BLam' \cap \Pi_2 \BLam'' = \emptyset$.

Finally,  to reach the same conclusion for
$\Pi_1 \BLam' \cap \Pi_2 \BLam''$ and  $\Pi_2 \BLam' \cap \Pi_1 \BLam''$,
it suffices to replace $\BLam'$ by $\sigma \BLam'$ and to use the
definition of $L$-D boxes:
$$
\min ( \dist(\BLam', \BLam''), \dist(\sigma \BLam', \BLam'')) > 8L.$$
Indeed, we have
$$\Pi_1 (\sigma \BLam') = \Pi_2 \BLam', \;
\Pi_2 (\sigma \BLam') = \Pi_1 \BLam'$$
so that an analogue of inequality (6.5) for boxes
$\sigma \BLam'$ and $\BLam''$ reads
$$\dist( \Pi_j (\sigma \BLam'), \Pi_j \BLam'') > 2L >0,
\eqno(6.6)$$
yielding
$$\dist( \Pi_2 \BLam', \Pi_1 \BLam'') > 2L >0,
\; \dist( \Pi_1 \BLam', \Pi_2 \BLam'') > 2L >0,
\eqno(6.7)$$
so that $\Pi_2 \BLam' \cap \Pi_1 \BLam'' = \emptyset$,
$\Pi_1 \BLam' \cap \Pi_2 \BLam'' = \emptyset$. Now we see that
$$\left(\Pi_1 \BLam' \cup \Pi_2 \BLam'\right) \cap
\left(\Pi_1 \BLam'' \cup \Pi_2 \BLam''\right) = \emptyset.
\eqno(6.8)$$
This completes the proof of Lemma \ref{DistDiag}. $\QED$
\pmn

\pmn{\bf  Acknowledgments.} We thank the referees for
numerous suggestions improving the quality of the paper.
VC thanks The Isaac Newton Institute and DPMMS, University of
Cambridge, for hospitality during visits in 2003, 2004,
2007 and 2008. YS thanks the D\'{e}partement de Math\'{e}matiques,
Universit\'{e} de Reims for hospitality during visits in
2003 and 2006--2008, in particular, for a Visiting Professorship
in the Spring of 2003. YS thanks IHES, Bures-sur-Yvette, and
STP, Dublin Institute for Advanced Studies, for hospitality
during visits in 2003--2007. YS thanks the Departments of Mathematics
of Penn State University  and of UC Davis, for hospitality
during Visiting Professorships in the Spring of 2004, Fall of
2005 and Winter of 2008.  YS thanks the Department of Physics,
Princeton University and the Department of Mathematics of UC Irvine,
for hospitality during visits in the Spring of 2008. YS
acknowledges the support provided by the ESF Research Programme
RDSES towards research trips in 2003--2006.

\bibliographystyle{amsalpha}

\end{document}